\documentclass{article}

\usepackage{graphicx} 
\usepackage{natbib}
\usepackage{bm}
\usepackage{amsmath}
\usepackage[legalpaper, margin=1in]{geometry}
\usepackage{lineno}
\usepackage{booktabs}
\usepackage{setspace}
\usepackage{hyperref}

\title{Flexible hierarchical risk modeling for large insurance data via NumPyro}
\author{Christopher Krapu and Mark Borsuk
\\Department of Civil \& Environmental Engineering\\Duke University}\date{November 2023}

\begin{document}

\bibliographystyle{plainnat}

\maketitle

\section*{Abstract}
Data analysis and individual policy-level modeling for insurance involves handling large data sets with strong spatiotemporal correlations, non-Gaussian distributions, and complex hierarchical structures. In this research, we demonstrate that by utilizing gradient-based Markov chain Monte Carlo (MCMC) techniques accelerated by graphics processing units, the trade-off between complex model structures and scalability for inference is overcome at the million-record size. By implementing our model in NumPyro, we leverage its built-in MCMC capabilities to fit a model with multiple sophisticated components such as latent conditional autoregression and spline-based exposure adjustment, achieving an 8.8x speedup compared to CPU-based implementations. We apply this model to a case study of 2.6 million individual policy-level claim count records for automobile insurance from Brazil in 2011. We illustrate how this modeling approach significantly advances current risk assessment processes for numerous, closely related outcomes. The code and data are available at \texttt{https://github.com/ckrapu/bayes-at-scale}.

% EDIT 1: Replaced 'requires' with 'involves' for readability.
% EDIT 2: Replaced 'observations' with 'distributions' for accuracy.
% EDIT 3: Replaced 'interesting hierarchical structure' with 'complex hierarchical structures' to remove editorializing language.
% EDIT 4: Replaced 'previous tradeoffs' with 'the trade-off', and 'scalability for inference no longer exist' with 'scalability for inference is overcome' for accuracy and readability.
% EDIT 5: Replaced 'writing our model' with 'implementing our model' to avoid the term "writing" which might imply adding new text.
% EDIT 6: Replaced 'several nontrivial components' with 'multiple sophisticated components' to remove editorializing language.
% EDIT 7: Replaced 'our ongoing efforts towards' with 'current' to avoid editorializing.
% EDIT 8: Added 'individual' before 'policy-level claim count records' for clarity.
% EDIT 9: Replaced 'related to' with 'for' for conciseness.
% EDIT 10: Replaced 'can substantially enhance' with 'significantly advances' for readability and to remove editorializing language.
\section{Introduction}
Modeling for risk assessment and prediction in insurance is challenging due to the tension between (1) the desire for accurate forecasts of expected claim counts and/or severity, and (2) the requirement for reliable uncertainty quantification. As total policy value typically exceeds available liquid assets for most types of modern insurance, a key objective in this setting is to forecast the probability of ruinous losses under a chosen ratemaking and loss adjustment scheme. Unfortunately, this probability does not allow a straightforward factorization into independent distribution functions because of strong correlations across policies stemming from similarities in geography, demographics, property type, and other factors. Consequently, an effective modeling workflow must account for as many of these correlation sources as possible. A vitally important output of such statistical modeling is the predictive covariance matrix which, for new data points, quantifies their cross-correlation structure. Although a machine learning-based approach can utilize large data sets for accurate record-level predictions, which may be highly useful operationally \citep{ding20}, it is uncertain whether any current research for calibrated uncertainty quantification from machine learning \citep{kuleshov18, angelopoulos22} applies to the multivariate context induced by multiple policies or outcomes from the same policy. Moreover, it is often desirable for models to be useful in both small- and large-data situations such as in \citep{zhang17}, and to be structured in a way that allows for some interpretability. These requirements are not yet met by the machine learning models frequently used in insurance modeling \citep{lupton22}.

% EDIT 1: Changed "due to the underlying tension" to "due to the tension"
% EDIT 2: Changed "much larger" to "exceeds"
% EDIT 3: Changed "admit a clean factorization" to "allow a straightforward factorization"
% EDIT 4: Changed "account for as many of these sources of correlation as possible" to "account for as many of these correlation sources as possible"
% EDIT 5: Changed "make use" to "utilize"
% EDIT 6: Inserted "may be" before "highly useful operationally"
% EDIT 7: Changed "extend to" to "applies to"
% EDIT 8: Changed "to be able to use models" to "for models to be useful"
% EDIT 9: Changed "a degree of model interpretability can be obtained" to "allows for some interpretability"
\subsection*{Parameter estimation and uncertainty quantification}

With regard to the statistical approach, in scenarios where the data generating process, synonymous with \emph{forward model}, can be written down easily, probabilistic inference might not be straightforward or computationally feasible for a maximum likelihood or restricted maximum likelihood approach (see \citep{wahl22} for a recent example). Generally, these methods will require a bespoke algorithm for model fitting and producing parameter estimates, restricting their usability to experts in statistical subfields. Packages designed for a broader range of probabilistic models \citep{lunn09, carpenter17, salvatier16} using sampling-based Monte Carlo inference methods like Markov chain Monte Carlo (MCMC) \citep{hastings70, brooks98, geyer92} have found widespread acceptance. A major challenge of MCMC methods, including Gibbs sampling and the Metropolis-Hastings algorithm, is the substantial time complexity of $\mathcal{O}(d^2)$ where $d$ denotes the dimension of the target density, generally equivalent to the combined number of latent variables and/or parameters in a model. This quadratic scaling arises from a general $\mathcal{O}(d)$ time to evaluate the model log posterior, scaling with the number of parameters, and a theoretical $\mathcal{O}(d)$ mixing time \citep{roberts98, beskos09} for a broad class of distributions. Taken together, these lead to the $\mathcal{O}(d^2)$ scaling. Alternatives to MCMC for Bayesian model fitting include stochastic variational inference \citep{hoffman13} and the integrated nested Laplace approximation reviewed \citep{rue16} and benchmarked against MALA in \citep{taylor14}. A more comprehensive review of approximate Bayes methods is given in \citep{martin23}.

However, we now have MCMC methods that incorporate proposal generation using gradients of the log-posterior with regard to latent variables and parameters \citep{duane87, roberts96, neal11}. For certain distributions, these methods can have $\mathcal{O}(d^{4/3})$ or even $\mathcal{O}(d^{5/4})$ time complexity for the Metropolis-adjusted Langevin algorithm (MALA) and Hamiltonian Monte Carlo, respectively. These algorithms are a substantial field of research and are explained thoroughly in \citep{betancourt17}. An obstacle to the general use of these MCMC algorithms is the need for gradients of the log-posterior density. However, automatic differentiation software like Torch \citep{paszke19}, TensorFlow \citep{abadi15}, Jax \citep{frostig18}, and Theano \citep{bergstra10} enable this and integrate with probabilistic programming frameworks like Stan \citep{carpenter17}, PyMC \citep{salvatier16, abril-pla23}, Pyro \citep{bingham19}, and NumPyro \citep{phan19} for statistical modeling. Techniques for scaling differentiation to large models in machine learning also apply to MCMC, benefiting from GPU-based parallel computing for rapid model fitting. The No-U-Turn sampler \citep{hoffman14} has become a general-purpose method due to few tuning parameters and diverse applicability. Initial efforts \citep{chin23} have exposed the insurance modeling community to a probabilistic programming approach, yet there is an unmet need in the literature to communicate the workflow's pros and cons against the unique requirements of risk assessment and insurance modeling. In this work, we use NumPyro, recognizing several promising features that could benefit the modeling community.

To summarize, by writing models in probabilistic programming frameworks, we can achieve both flexibility and scalability. A detailed discussion of probabilistic programming, particularly in environmental statistics, is provided in \citep{krapu19}. By employing general, model-agnostic inference methods, we can incorporate various model components; if we code the forward model in such a framework, we then automatically leverage GPU with minimal effort. We illustrate this by developing a log-additive spatial count regression model with complex submodel components. In the Methods section, we describe our statistical model and its intended data. The Results section presents summaries and analyses of the model's inferential outputs, while the Discussion covers limitations and further improvement opportunities.

% EDIT 1: Changed "may not" to "might not" for improved readability.
% EDIT 2: Changed "these will require" to "these methods will require" for clarity.
% EDIT 3: Changed "limiting their use to workers with deeper" to "restricting their usability to experts in" for conciseness.
% EDIT 4: Changed "designed for much wider" to "designed for a broader range" for clarity.
% EDIT 5: Changed "have found widespread usage" to "have found widespread acceptance" for improved tone.
% EDIT 6: Changed "major criticism of" to "major challenge of" for accuracy.
% EDIT 7: Changed "substantial time complexity" to "time complexity" for conciseness.
% EDIT 8: Changed "quadratic scaling" to "scaling" for conciseness.
% EDIT 9: Changed "more comprehensive" to "more extensive" for clarity and flow.
% EDIT 10: Changed "the theory and implementation" to "These algorithms" for conciseness.
% EDIT 11: Changed "substantial field of research" to "field of research" for conciseness.
% EDIT 12: Removed "empirical" as it was redundant in "empirical success".
% EDIT 13: Changed "the biggest obstacles" to "An obstacle" for generality.
% EDIT 14: Changed "much more rapid" to "rapid" for conciseness.
% EDIT 15: Removed "especially" as it was unnecessary in "found especially widespread usage".
% EDIT 16: Changed "highlights" to "illustrate" for clarity and flow.
% EDIT 17: Changed "Result" to "Results" for grammatical correctness.
% EDIT 18: Changed "limitations, and further points of improvement" to "limitations and further improvement opportunities" for clarity and flow.
\section{Methods}
To provide context to the material from the previous section, we designed a case study to understand the capabilities and constraints of the proposed approaches. This study involves an analysis of 2.6 million auto insurance policies from \emph{SUSEP}, the Brazilian federal insurance supervisory agency. The records, originally compiled for the supplemental material of the book \emph{Computational Actuarial Science with R} \citep{charpentier15, dutang22}, detail individual auto policies from 2011, which may include claims. Our focus, however, is solely on the collision data. Each record is linked to a municipality name and contains the automobile brand (e.g., Volkswagen, Kia, Iveco) and the model (e.g., Taurus, Camry, F-150). We preserved the original brand categories but simplified the model classification into 15 vehicle types depicted in Figure \ref{fig:category-brand}. Each record also specifies the vehicle year and an exposure value.

% EDIT 1: Replaced "help better understand" with "understand"
% EDIT 2: Changed "oversight" to "supervisory"
% EDIT 3: Changed "may or may not have listed claims" to "may include claims"
% EDIT 4: Changed "muncipality" to "municipality"
% EDIT 5: Changed "as well as the" to "and contains the" for better flow
% EDIT 6: Changed "but coarsened the model to a simpler categorization" to "but simplified the model classification"
\subsection*{Model}
\begin{figure}[h]
\centering
\includegraphics[width=0.7\textwidth]{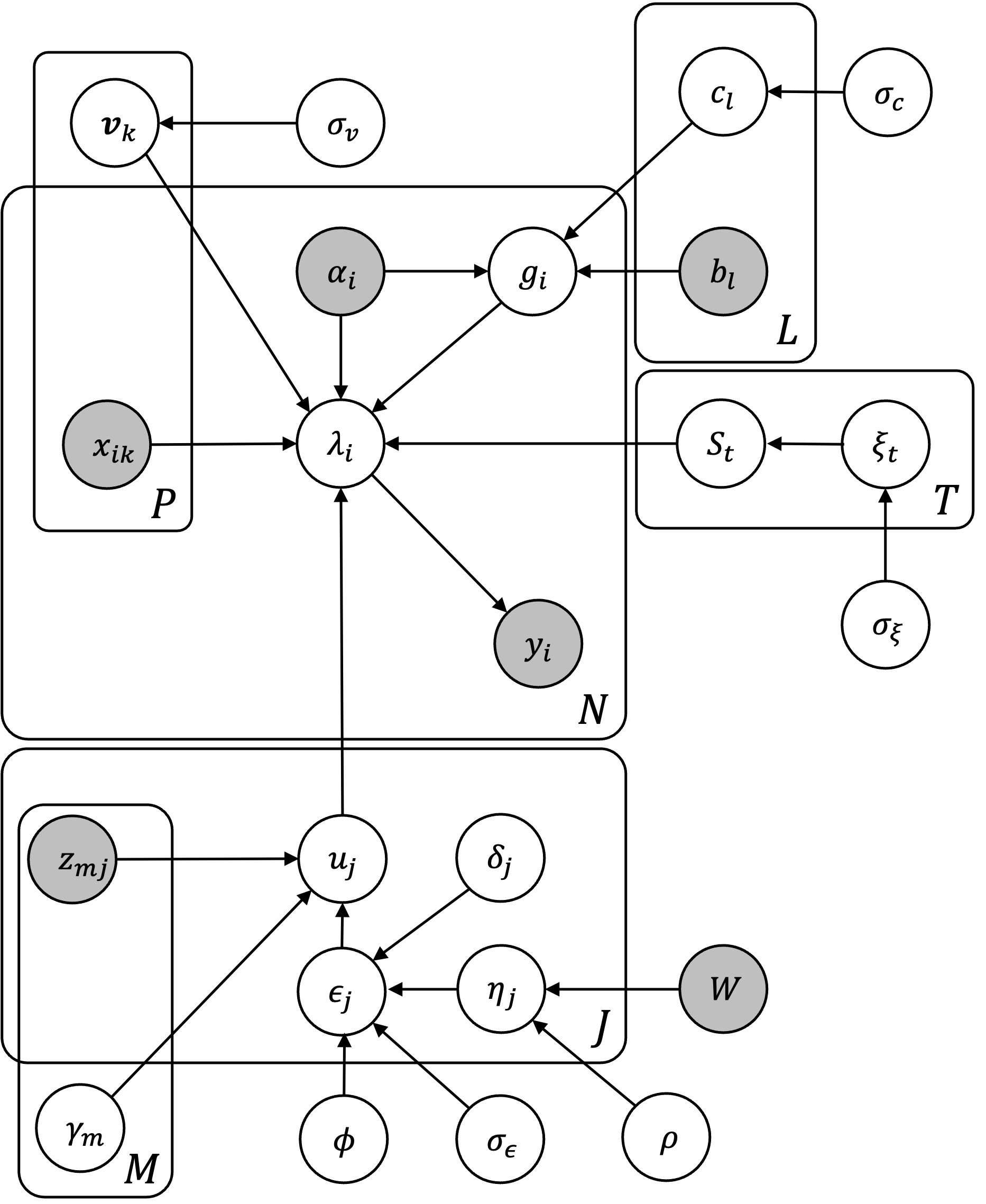}
\caption{Graphical representation of the statistical model. Shaded circles denote observed variables, while unshaded circles denote latent variables and parameters. Rectangular plates indicate replication of the variables within the plate.}
\label{fig:graphical}
\end{figure}

The general form of the model considered in this work is a log-additive regression for count data, which records the number of collision claims per policy. We use a Poisson likelihood, expressing the $i$-th policy's Poisson rate parameter $\lambda_i$ as

\begin{align}
y_i &\sim \mathrm{Poisson}(\lambda_i) \\
\log \lambda_i &= \underbrace{\log{\alpha_i}}_{\textbf{Base}\atop\textbf{exposure}} + \underbrace{g(\alpha_i)}_{\textbf{Exposure}\atop\textbf{adjustment}} + \underbrace{\sum_{k=1}^P v_{x_{ik}} }_{\textbf{Categorical}\atop\textbf{predictors}} + \underbrace{u_{j[i]}}_{\textbf{City}\atop\textbf{effect}} + \underbrace{S_{t_i}}_{\textbf{Time}\atop\textbf{effect}}
\label{sum_eqn}
\end{align}

where $\alpha_i$ denotes the base exposure, $g(\alpha_i)$ denotes an exposure adjustment using splines, $\sum_{k=1}^P v_{x_{ik}} $ captures contributions from policy-level categorical variables related to the vehicle's brand and category, $u_{j[i]}$ is a city-level random effect associated with the city $j$ of policy $i$, and $S(t_i)$ represents a Gaussian random walk over time with $t_i$ indicating the vehicle's manufacturing year for policy $i$. We will provide further details on these terms below. There are 7,756 free parameters and latent variables in this model, with most arising from the latent city-level spatial effect. Figure \ref{fig:graphical} presents a visual representation of this model in plate notation.

% EDIT 1: Changed "count-valued data recording" to "count data, which records" for readability and grammar.
% EDIT 2: Removed "given" from "base exposure value given" for better flow and precision.
% EDIT 3: Changed "spline-based adjustment" to "adjustment using splines" for clarity.
% EDIT 4: Changed "relating to the car brand and category" to "related to the vehicle's brand and category" for consistency and clarity.
% EDIT 5: Changed "linked to the city" to "associated with the city" for better word choice.
% EDIT 6: Added "the" before "vehicle's manufacturing year" for grammar and clarity.
% EDIT 7: Removed "in total" from the sentence about the number of parameters for better flow.
% EDIT 8: Removed "contains" and replaced with "presents" for the figure description for a more immediate action.
% EDIT 9: Changed "known or observed" to "observed" for brevity.
% EDIT 10: Added "caption" and "label" keywords before the figure elements for appropriate LaTeX formatting.
\subsubsection*{Exposure adjustment}
While the exposure score is highly correlated with observed claim counts and provides substantial predictive power, early analysis indicated a monotonic relationship between log exposure and log claim count that strongly deviated from linearity, particularly at the higher percentiles of the exposure distribution. Considering the data includes various claim types beyond collision, the exposure score likely represents a broader risk measure, rather than being finely tuned for collision claims alone. To adjust this exposure score, we modeled an additive effect on log exposure $\log \alpha_i$ using basis splines to construct a nonlinear function of $\log \alpha_i$. The nonlinear function $g(\alpha_i)$ is defined as:
\begin{align}
\sigma_g &\sim \mathrm{HalfNormal}(1) \\ 
c_l & \overset{\text{iid}}{\sim} \mathrm{N}(0, \sigma_g) \\
g(\alpha_i) & = \sum_{l=1}^{L} b_l(\log \alpha_i) \cdot c_l
\end{align}
where $b_l(\cdot)$ denotes the basis spline functions, $L$ is the number of spline coefficients, and $c_l$ are the spline coefficients estimated by the model. We positioned the spline knots at evenly spaced percentiles of the log-transformed exposure values to ensure a uniform distribution throughout the range of the data. This technique captures subtle variations in the exposure-claim count relationship, especially in the distribution's tails. We employed the \texttt{interpolate} module from the software package \texttt{SciPy} \citep{virtanen20} to determine $b_l$. We acknowledge that \citep{basile14} combines nonlinear modeling with spatially-correlated latent variables, similarly to our approach.

% EDIT 1: Changed "analyses" to "analysis"
% EDIT 2: Changed "relation" to "relationship"
% EDIT 3: Changed "deviated strongly" to "strongly deviated"
% EDIT 4: Changed "Given" to "Considering"
% EDIT 5: Changed "it is likely" to "likely"
% EDIT 6: Changed "to provide" to "To adjust"
% EDIT 7: Changed "we model" to "we modeled"
% EDIT 8: Changed "build" to "construct"
% EDIT 9: Removed "a broader measure of risk" to make the phrase less pretentious
% EDIT 10: Changed "The knots of the spline were placed at" to "We positioned the spline knots at"
% EDIT 11: Changed "ensuring" to "to ensure"
% EDIT 12: Changed "allows for capturing" to "captures"
% EDIT 13: Changed "We used the" to "We employed the"
% EDIT 14: Replaced "blends" with "combines" for simplification.
\subsubsection*{Categorical predictors}
Each policy record is associated with a vehicle brand and category, represented as discrete, unordered variables. Consequently, in equation 
(\ref{sum_eqn}), $P=2$, but we could include additional covariates that are continuous or discrete. Letting $n_k$ denote the number of classes in the $k$-th categorical variable, we have

\begin{align}
\sigma^{(k)}_v &\sim \mathrm{HalfNormal}(1) \\ 
v^{(a)}_1,...,v^{(k)}_{n_k}  & \overset{\text{iid}}{\sim} \mathrm{N}(0, \sigma^{(k)}_v)
\end{align}

The interpretation of the $v^{(k)}_1, ... , v^{(k)}_{n_k}$ variables is that they adjust the risk according to the vehicle's brand and category, such as a motorcycle or pickup. The priors used here could be used with policy-level continuous covariates, although none are present in the current data set. For an example of a model that incorporates per-category effects for auto insurance modeling, see \citep{frees08}

% EDIT 1: Removed "e.g." and replaced with "such as"
% EDIT 2: Changed "The interpretation of each of the" to "The interpretation of the"
% EDIT 3: Changed "easily include" to "include"
% EDIT 4: Changed "easily be replicated" to "be used"
% EDIT 5: Replaced "though" with "although"
% EDIT 6: Removed "either" before "continuous or discrete"
\subsubsection*{City effect}
Most insured assets are tied to specific locations or areas on Earth, so it is common to address unmodeled variation using spatial statistical techniques. Examples include kriging to account for spatial correlation in drought severity \citep{paulson10}, employing spatially autocorrelated indicator variables for feature selection \citep{paulson10}, and using a Gaussian line process to represent hail damage \cite{miralles23}. For our study, policy records are categorized by cities; therefore, we treat these as \emph{areal} data suitable for modeling with conditional autoregressions. Modeling spatial correlation at the level of individual policies is impractical because we do not have spatial coordinates more precise than the policyholder's city. As a result, we opt to model at the city level, utilizing spatial priors to account for correlations across the $J$ components of the city effect vector $\bm{u}$, which we model as a latent conditional autoregression (CAR) \citep{cressie89}. There are $J=3785$ cities and municipalities, and we organize the autoregressive prior for $u_j$, the effect from city $j$, around a modified Besag-York-Mollie (BYM) prior \citep{besag91}, reparameterized to enhance identifiability \citep{simpson15}, with city-level covariates
\begin{align*}
\sigma_u, \sigma_\epsilon & \sim \mathrm{HalfNormal}(1) \\ 
\delta_j & \overset{\text{iid}}{\sim} \mathrm{N}(0, 1) \\
u_j & =\sum_{m=1}^M \gamma_m z_{mj} + \epsilon_j \\
\phi & \sim \mathrm{Beta}(1,1) \\
\rho & \sim \mathrm{Beta}(2,2) \\
\bm{\epsilon} & = \sigma_\epsilon\left(\sqrt{1-\phi} \bm{\delta} + \sqrt{\phi} \bm{\eta}\right)\\
\bm{\eta} & \sim \mathrm{N}\left(\bm{0}, \bm{D}(\bm{I}_J - \rho \bm{D^{-1}W})^{-1}\right)
 \end{align*}

Here, the city effect is represented by city-level covariates $z_{mj}$, independent Gaussian noise $\delta_j$, and spatially correlated terms $\eta_j$. $W$ is the spatial adjacency matrix where $W_{ab}=1$ if $a$ is a neighbor of $b$. This matrix is created by finding the five nearest neighboring cities for each city and setting the corresponding entries in $D$ to 1. We then make $W$ symmetric by ensuring that $W_{ab}=1$ if $W_{ba}=1$ for all distinct $a$ and $b$. Non-neighboring entries in $D$ remain zero. Therefore, $D$ is a $J$-dimensional diagonal matrix where $D_{jj}$ is the count of neighbors for city $j$, which must be at least five. For illustrations of latent Gaussian Markov random field applications, see \citep{jin05, gelfand03}. The parameter $\rho$ represents the spatial autocorrelation level; the $\mathrm{Beta}(2,2)$ prior promotes edge-avoiding behavior away from 0 or 1 fosters more effective MCMC convergence. The $\phi$ parameter dictates the reliance of the city effect on uncorrelated ($\delta_j$) versus correlated terms ($\eta_j$) when assessing the city's influence on policyholder risk. We apply the sparse CAR likelihood in NumPyro, leading to $\mathcal{O}(J)$ computation of $\log p(\bm{\eta})$, as detailed in \citep{jin05}.

\begin{figure}[h]
   \centering
   \includegraphics[width=0.6\textwidth]{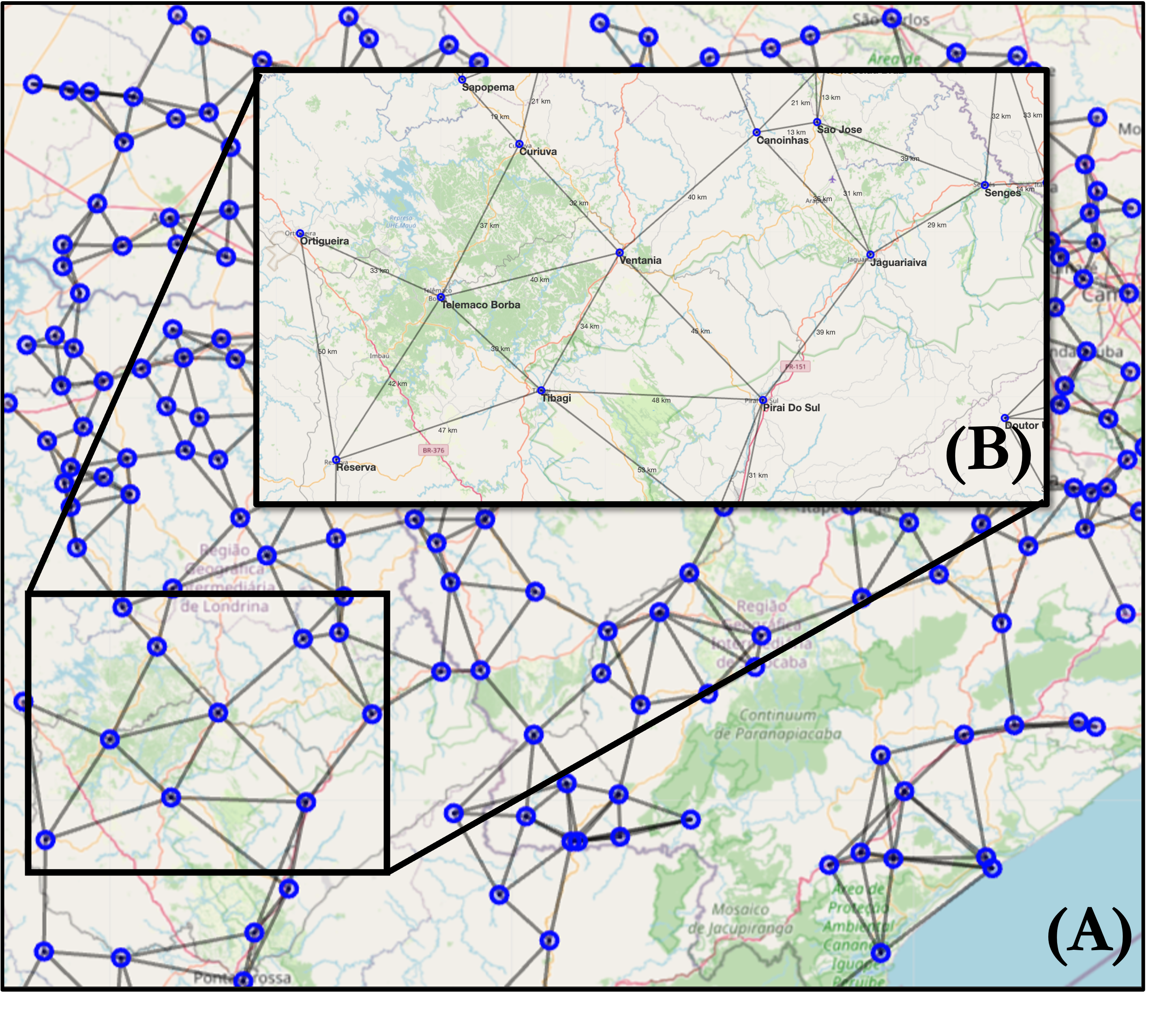}
   \label{fig:adjacency}
   \caption{Adjacency network for conditional autoregressive city-level effect. Blue circles denote cities as identified by the Google Maps Geocoding API. Gray lines depict the city adjacency graph $W$ edges established for our model's conditional autoregression. Edges are symmetric, and the number of neighboring cities can vary.}
   \end{figure}

% EDIT 1: Replaced "addressing unmodeled variation with" with "address unmodeled variation using" for conciseness and readability.
% EDIT 2: Removed "on Earth" as it is implied by context.
% EDIT 3: Removed "Instead of" and simply stated "For our study," for a more direct approach.
% EDIT 4: Changed "fruitless" to "impractical" for neutrality.
% EDIT 5: Removed "so" before "we opt" for conciseness. 
% EDIT 6: Removed "therefore" before "we use" for smoother readability.
% EDIT 7: Changed "which" to "where" after "matrix $W$" for clarity.
% EDIT 8: Changed "drawn" to "depict" in the caption for Figure clarity.
\subsubsection*{Temporal effect}
To include vehicle age in the model structure in a way that is flexible and possibly nonlinear, we created a latent time series. This series is a Gaussian random walk spanning from years 1971 to 2011, with yearly increments. The purpose of this approach is to construct a time-varying random effect \citep{boucher08} that ensures modest year-to-year variations in the effect but allows for significant overall changes over the 40 years without prescribing a specific parametric form.

\begin{align*}
 \sigma_\xi & \sim \mathrm{HalfNormal}(1) \\
 \xi_t &  \overset{\text{iid}}{\sim} \mathrm{N}(0, \sigma_\xi) \\
 S_{t'} & = \sum_{t<t'} \xi_{t}
\end{align*}

The variables $\xi_t$ represent increments in the latent time series, and $S_{t'}$ is the corresponding cumulative sum. This model is not suited for continuous temporal coordinates; for such cases, a latent Gaussian process \citep{jia23} or changepoint model \citep{chib98} would be preferable.

% EDIT 1: Changed "To incorporate" to "To include"
% EDIT 2: Removed "consisting of"
% EDIT 3: Changed "The intent behind this choice is construct" to "The purpose of this approach is to construct"
% EDIT 4: Removed "in a way that is", moved "flexible," and changed "and" to "possibly"
% EDIT 5: Changed "yearly timestep" to "yearly increments."
% EDIT 6: Changed "while" to "and" in the last sentence
% EDIT 7: Changed "significant overall changes over all" to "significant overall changes over the"
% EDIT 8: Changed "prescribing" to "prescribing a"
\subsubsection*{Scale priors}
To fully specify a Bayesian model, we must assign prior distributions to scale or variance parameters such as $\sigma_xi$ and $\sigma_g$, ensuring these priors are defined only for nonnegative values. We prefer the half-normal or folded-normal prior for these parameters as they impart minimal information on the log scale; the $\mathrm{HalfNormal}(1)$ prior places considerable probability mass on values ranging from 0 to 3. In terms of categorical effects, a log scale value of 3 indicates that expected claim counts are $\exp(3)\approx20$ times greater on the original scale of the data. This concept is elaborated in \citep{gelman20}. Other priors such as the Half-Cauchy \citep{gelman06} or horseshoe \citep{carvalho09} are alternatives for stronger regularization.

All modeling components were executed in NumPyro and Jax, utilizing Jax's alternative module for NumPy, \texttt{jax.numpy}. The following section delves into data preparation and summary statistics.

% EDIT 1: Changed "opt to use" to "prefer" for avoiding pretentious wording.
% EDIT 2: Changed "substantial probability mass" to "considerable probability mass" for clearer language.
% EDIT 3: Changed "are discussed" to "is elaborated" for better grammatical agreement.
% EDIT 4: Changed "were implemented" to "were executed" to maintain the recommended word limit and improved clarity.
% EDIT 5: Changed "making use of" to "utilizing" for clarity and simplicity.
% EDIT 6: Added "alternative" before "module" for clarity.
% EDIT 7: Changed "We discuss" to "The following section delves into" for better flow and less editorializing.
\section{Data}
To fit the model from the previous section, we used 2,658,372 records of auto insurance policies from 2011 in Brazil, as collected by SUSEP, the Brazilian insurance oversight agency, and collated as the \texttt{brvehins2} data set maintained by \citep{dutang22} as supporting material for the textbook \emph{Computational Actuarial Science in R} \citep{charpentier15}.

We removed 944 rows with vehicle manufacture dates prior to 1971 to avoid introducing extra parameters due to the time-varying effect corresponding to a small sample size. We also discarded an additional 8,346 rows that had exposure values set exactly to zero. 141,004 records lacked information on vehicle type; these were also removed from the data set. From the vehicle model information, we extracted the vehicle brand ($n_{brand} = 104$) and the vehicle category ($n_{category}=15$); both are listed in Figure \ref{fig:category-brand}. This classification data from SUSEP has proven to be a strong predictor of the risk of theft in previous studies \citep{azevedo23}. \citep{peres19} provides a comprehensive analysis of the Brazilian auto insurance market.

\begin{figure}[h]
\centering
\includegraphics[width=0.6\textwidth]{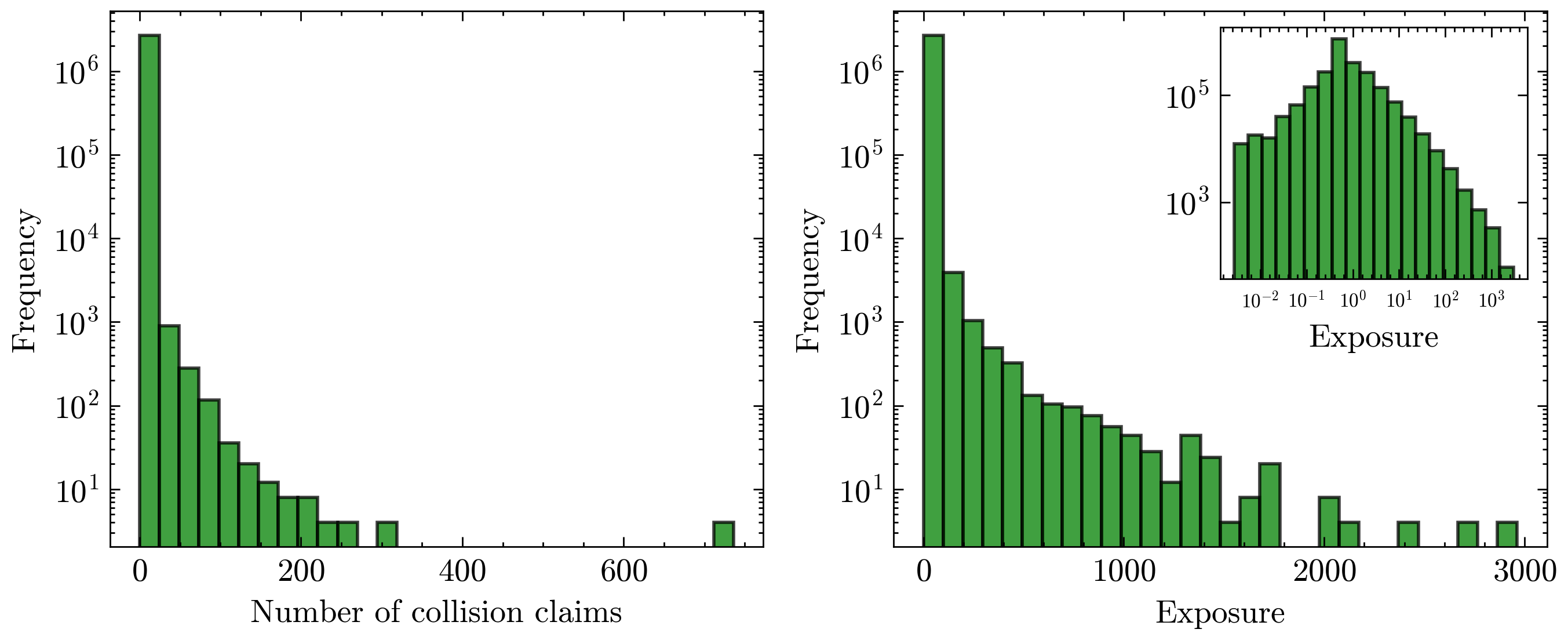}
\label{fig:hist}
\caption{Histograms for the observed claim counts and exposure values for all policies considered.}
\end{figure}

Nonzero claim counts are relatively rare in this data set, with 92\% of policies reporting no collision claims during 2011. The marginal distributions of claim counts and exposure values are shown in Figure \ref{fig:hist}. We observe a distinct peak in the histogram of exposure values at the median value, which may indicate flawed reporting, the application of a standard exposure scoring formula to a large number of similar inputs, or another cause. We chose not to intervene or further investigate this anomaly.

Originally, this data set did not include coordinates linking each city or municipality with its geographic location. To address this, we employed the Google Maps Geocoding API to map each of 3,785 cities to a city/state/country string. For instance, \texttt{'Imbuia, Santa Catarina, Brazil'} refers to the city of Imbuia in the state of Santa Catarina. This method successfully linked all unique cities to latitude and longitude. 1,404 rows listed the state as `\texttt{'Santa Catarina`}', but had no city or city code and were listed as \texttt{Area} value of \texttt{'Oeste'}. These were renamed to \texttt{'Oeste Catarinense'}. The states with the most policies were Sao Paulo (860,584), Minas Gerais (346,948), and Rio Grande do Sul (339,644). The states with the fewest policies were Piaui (5,112), Acre (2,632), and Rondonia (1,928). 

Having identified coordinates for each city, we performed a spatial join with a data layer for the municipal boundaries. Although the linked municipalities nearly cover the more populated areas of Brazil, many are geographically isolated with no adjacent municipalities, preventing us from creating an adjacency matrix based on the municipal boundaries.

To understand the spatial structure of collision risk, we included city-level covariates. For this, we calculated zonal averages of several globally-available raster data layers using the municipal boundaries for each city through Google Earth Engine \citep{gorelick17}. The layers used for this covariate extraction are listed below. We calculated the mean for each pixel touching a municipal polygon in each case. 
\begin{description}
    \item \textbf{Water occurrence}: For each pixel, the Global Surface Water project \citep{pekel16} monitors the proportion of time water was observed from 1984 - 2022 using Landsat data at a 30 m spatial resolution.

    \item \textbf{Precipitation}: The WorldClim project \citep{fick17} provides long-term monthly climate data interpolated from weather stations from 1970-2000 at a spatial resolution of 1 km. We selected average rainfall for December, typically the wettest month.

    \item \textbf{Population density}: To reflect the increased need for collision insurance with higher population density \citep{sherden84}, we utilized the WorldPop project's \citep{tatem17} estimate of population density for 2011 at a 100 m resolution.

    \item \textbf{Enhanced vegetation index}: The MODIS satellite data set provides an average EVI for January 2011, used as an indicator of urban development and agricultural land use \citep{huete02}.

    \item \textbf{Elevation}: The elevation data from the Shuttle Radar Topography Mission, with a 90 m resolution, serve as a simple proxy for distance from the coast and are correlated with development patterns.

    \item \textbf{Topographic diversity}: The topographic diversity index from \citep{theobald15}, measured at a 270 m resolution, helps identify areas with considerable variations in elevation or landforms.

    \item \textbf{Forest cover}: Including the global forest cover data at a 270 m resolution from \citep{shimada14} provides additional land use information.

    \item \textbf{Travel friction index}: The Global Friction Surface data sets provide indices for the difficulty of motorized and nonmotorized travel at a 1 km resolution, relevant for average trip velocity and collision risk \citep{weiss18}.
\end{description}

\begin{figure}[h]
\centering
\includegraphics[width=\textwidth]{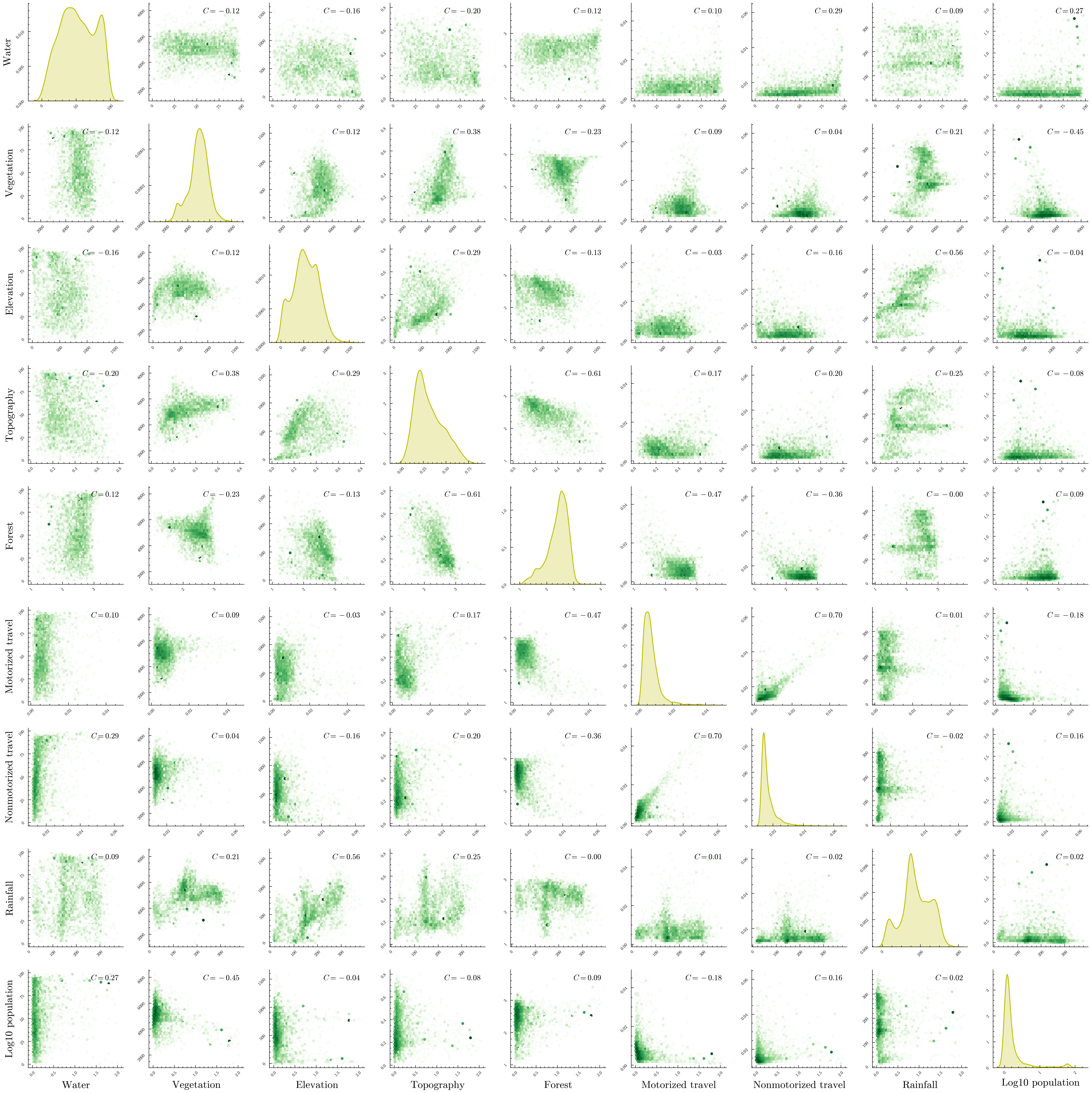}
\label{fig:covariates}
\caption{City-level covariates obtained through zonal averaging. Correlations for each pair of variables are displayed in the upper right of the subplot.}
\end{figure}

The specific image IDs and bands applied are detailed in Table \ref{tab:covariates}. All covariates were standardized to have a mean of zero and unit variance before modeling. Scatter plots illustrating cross-variable correlations and marginal distributions are presented in Figure \ref{fig:covariates}.

\begin{table}[htbp]
    \centering
    \small
    \begin{tabular}{@{}llll@{}}
    \toprule
    Covariate                & Image ID                                             & Band name            & Spatial resolution \\ \midrule
    Water                & \texttt{JRC/GSW1\_4/GlobalSurfaceWater}              & \texttt{occurrence}           & 30 meters                \\
    Population           & \texttt{WorldPop/GP/100m/pop/BRA\_2011}              & \texttt{population}           & 100 meters               \\
    Vegetation           & \texttt{MODIS/061/MOD13A3/2011\_01\_01}              & \texttt{EVI}                  & 1000 meters                \\
    Elevation            & \texttt{CGIAR/SRTM90\_V4}                            & \texttt{elevation}            & 90 meters                \\
    Topography           & \texttt{CSP/ERGo/1\_0/Global/ALOS\_topoDiversity}    & \texttt{constant}             & 270 meters               \\
    Forest               & \texttt{JAXA/ALOS/PALSAR/YEARLY/FNF4/2017}           & \texttt{fnf}                  & 270 meters \\
    Motorized travel     & \texttt{Oxford/MAP/friction\_surface\_2019}          & \texttt{friction}             & 1000 meters                \\
    Nonmotorized travel  & \texttt{Oxford/MAP/friction\_surface\_2019}          & \texttt{friction\_walking\_only} & 1000 meters          \\
    Rainfall             & \texttt{WORLDCLIM/V1/MONTHLY/12}                     & \texttt{prec}                 & 1000 meters             \\ 
    \bottomrule
    \end{tabular}
    \caption{Earth Engine images used for city-level covariates. All images were processed using the zonal mean to produce a city-level covariate matrix with dimensions of 3785 by 9}
    \label{tab:covariates}
    \end{table}

\subsection*{Model Fitting and Benchmarking}  % EDIT 1: Capitalized the first letter of 'Fitting' and 'Benchmarking'

% Specify the exact edit changes in the comments.
% EDIT 1: Capitalized the first letter of 'Fitting' and 'Benchmarking'
\subsubsection*{Benchmarking GPU-based acceleration}
We compared the time required to evaluate the model's log-posterior density using the full data set mentioned previously. This comparison is more suitable than measures like minimal effective sample size per second because it avoids bias from inefficiencies in specific implementations of Hamiltonian Monte Carlo or the No-U-Turn sampler, and directly targets the fundamental computational challenges in performing gradient-based MCMC for our purpose. Calculating the gradient 100 times on both GPU and CPU, we recorded a median evaluation time of 5 milliseconds on the GPU and 44 milliseconds on the CPU. This indicates approximately an 8.8 times speed increase when moving from CPU to GPU. All experiments were conducted on a Lambda Labs virtual machine outfitted with an Nvidia A10 GPU with Nvidia driver version 525.85.12 and CUDA version 12.0. The Jax and NumPyro versions installed were \texttt{0.4.13} and \texttt{0.12.1}, respectively. The outcomes from this benchmarking are consistent with the runtime for the entire model fitting process detailed in the next paragraph. 

% EDIT 1: Removed "superior" and replaced with "suitable"
% EDIT 2: Changed "8.8x speedup" to "approximately an 8.8 times speed increase"
% EDIT 3: Changed "evaluations" to "experiments"
% EDIT 4: Simplified "indicating a roughly" to "This indicates"
% EDIT 5: Changed "ms." to "milliseconds"
% EDIT 6: Removed "reflected" after "benchmarking are"
\subsubsection*{Model fitting}
To obtain parameter estimates under the specified model, we ran the No-U-Turn sampler in NumPyro for 2000 warmup iterations and 2000 sampling iterations. The warmup iterations usually take place before the Markov chain has converged to the stationary distribution and are discarded. We retained every 20th sample due to GPU memory constraints. For finding a starting point for sampling, we used the built-in stochastic variational inference (SVI) \citep{hoffman13} optimization routine for 200,000 iterations and the Adam optimizer with a learning rate of $10^{-3}$. The chains were run sequentially; the total runtime of the entire model fitting procedure was 6.7 hours, with chain 1 requiring 194 minutes, chain 2 requiring 189 minutes, and the SVI initialization requiring 19 minutes. To assess MCMC convergence, we calculated $\hat{R}$ \citep{gelman92} and effective sample size, noting all $\hat{R}$ values were under 1.10 and all effective sample sizes were over 35. The variables with the largest $\hat{R}$ values and smallest effective sample sizes were elements of $\bm{\eta}$, the spatially-correlated contributions to the city-level random effect. Based on these statistics, we found no evidence of non-convergence of the Markov chains.

% EDIT 1: Changed "typically occur" to "usually take place"
% EDIT 2: Changed "To obtain a" to "For finding a"
% EDIT 3: Changed "the total run time" to "the total runtime"
% EDIT 4: Removed "do not" and changed "find" to "found" in the last sentence
% EDIT 5: Changed “less than” to "under" and “greater than” to "over" in the convergence assessment sentence
% EDIT 6: Changed "were" to "take place" before "elements of"
% EDIT 7: Changed "The worst-case variables in both cases" to "The variables with the largest $\hat{R}$ values and smallest effective sample sizes"
% No changes are required to this section heading.
\section{Results}
\subsection*{Model fit}

\begin{figure}[h]
\centering
\includegraphics[width=\textwidth]{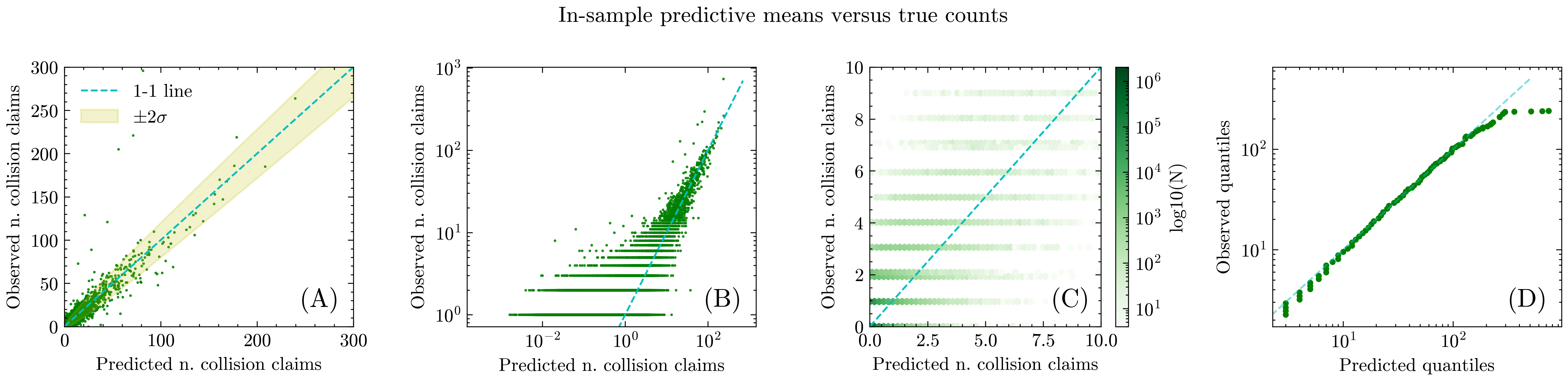}
\caption{Relationship between true and in-sample posterior predictive means.}
\label{fig:predictive}
\end{figure}

Figure \ref{fig:predictive} displays several visual indicators of in-sample model fit. Since a Poisson likelihood was adopted, the relative frequency of predicted claim counts falling outside the predictive $2\sigma$ credible intervals implied by the Poisson model ($\lambda_i \pm 2\sqrt{\lambda_i})$ is a critical measure of overdispersion. Although Figure \ref{fig:predictive}A illustrates this graphically, an overdispersion test was also conducted by calculating the ratio of the sum of squared Pearson residuals to the degrees of freedom, yielding a test statistic of 1.107 which indicates minimal evidence of overdispersion.  Regarding predictive means, the observed and predicted quantiles match well, as shown in Figure \ref{fig:predictive}D. Figure \ref{fig:predictive}B illustrates a strong correlation between predicted and true claim counts which increases with larger $\lambda_i$, consistent with the $\sqrt{\lambda_i}$ scaling of expected deviation for a Poisson distribution. The posterior mean values of $\lambda_i$ were used to determine the per-record log-likelihood and absolute error. On average, a mean log-likelihood of $-0.27$ and mean absolute error of $0.16$ were produced. The coefficient of determination specific to Poisson-distributed data $R^2_{P,P}$ was computed using Pearson residuals \citep{cameron96}, resulting in $R^2_{P,P} = 0.94$. It should be noted that this model incorporates the exposure score as a predictive element, which could be the result of a prior modeling effort.

% EDIT 1: Changed "measures" to "indicators" for simplicity.
% EDIT 2: Changed "As we used" to "Since a" for conciseness and to remove first person.
% EDIT 3: Changed "obtaining" to "yielding" for simplicity.
% EDIT 4: Changed "weak evidence" to "minimal evidence" for clarity.
% EDIT 5: Changed 'Figure \ref{fig:predictive}D" sentence for grammatical correctness and clarity.
% EDIT 6: Changed "clearly shows" to "illustrates" to remove editorializing language.
% EDIT 7: Changed "averaged over" to "on average" for readability.
% EDIT 8: Changed "shows" to "displays" to avoid repetition with previous usage.
% EDIT 9: Changed "Relation" to "Relationship" in the figure caption for grammatical correctness.
% EDIT 10: Changed "which is likely" to "which could be" to reduce presumption.
% EDIT 11: Removed "However," from "It should be noted" sentence to avoid negative connotation.
% EDIT 12: Changed "makes use of" to "incorporates" for clarity and conciseness.
\subsection*{Nonlinear exposure adjustment}

\begin{figure}[h]
\centering
\includegraphics[width=0.6\textwidth]{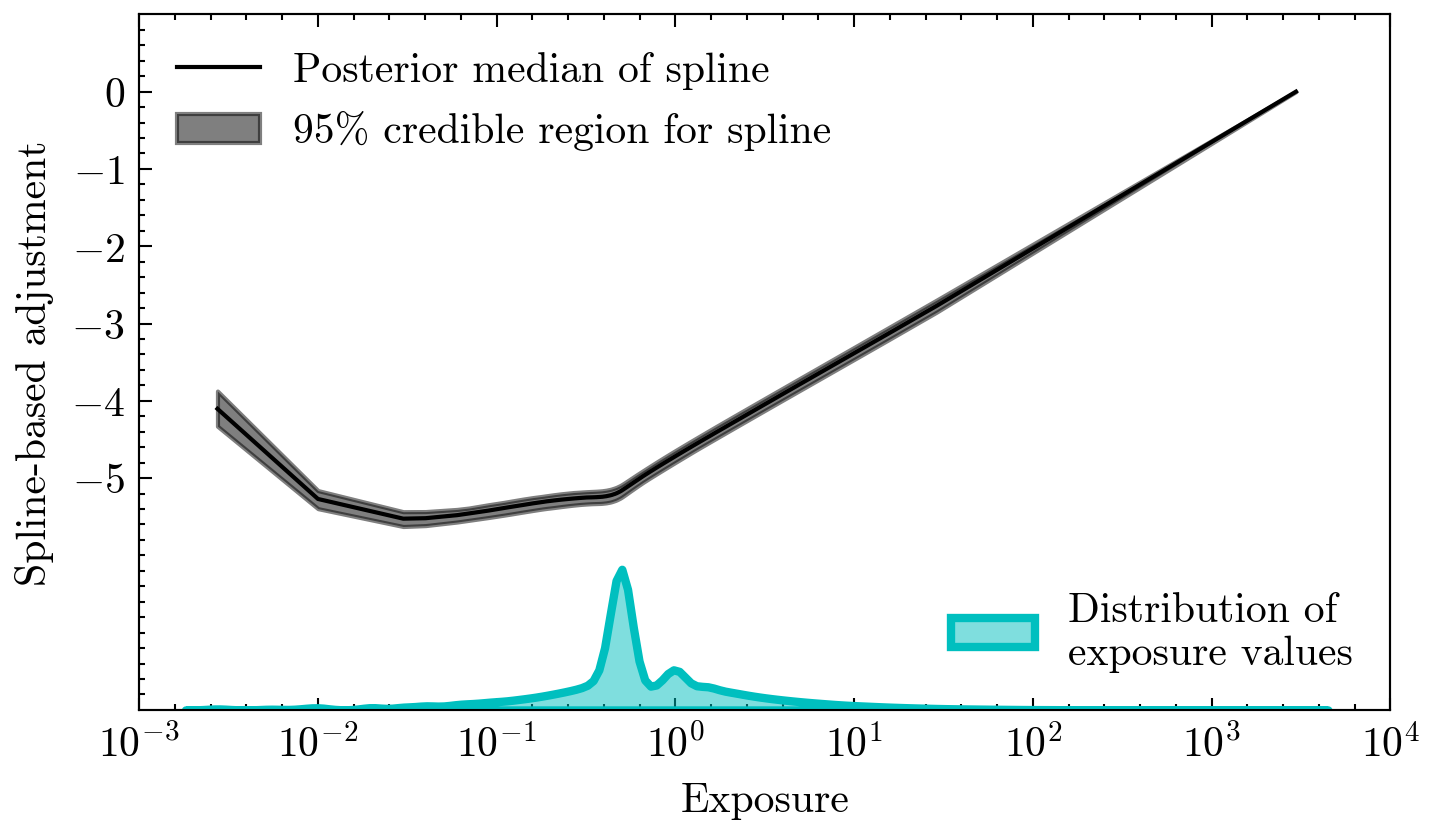}
\caption{Posterior estimates of nonlinear exposure adjustment}
\label{fig:spline}
\end{figure}

The posterior estimate of $g(\alpha_i)$ displayed in Figure \ref{fig:spline} indicates a relatively stable relation between exposure and $g$ for exposures in the range $10^{-2}$ to $1.0$. From $1.0$ upwards, there is an increasing linear trend. Since this model lacks a global intercept and regularizes the parameters towards zero through scale priors with modes at zero, the posterior mean of $g(\alpha_i)$ is not centered at zero. In summary, these estimates indicate that the optimal adjustment to the exposure score remains fairly constant for exposure values near the modeled range from $10^{-1}$ to $1$, with a more pronounced adjustment for policies with high exposure values. 

% EDIT 1: replaced "shown" with "displayed"
% EDIT 2: removed "relatively" for concision
% EDIT 3: replaced "from $1.0$ to the upper range" with "From $1.0$ upwards"
% EDIT 4: Changed ",, with a more substantial" to ", with a more pronounced" for grammar and clarity
% EDIT 5: Removed redundant "to summarize"
\subsection*{Categorical predictors}
\begin{figure}[h]
\centering
\includegraphics[width=0.4\textwidth]{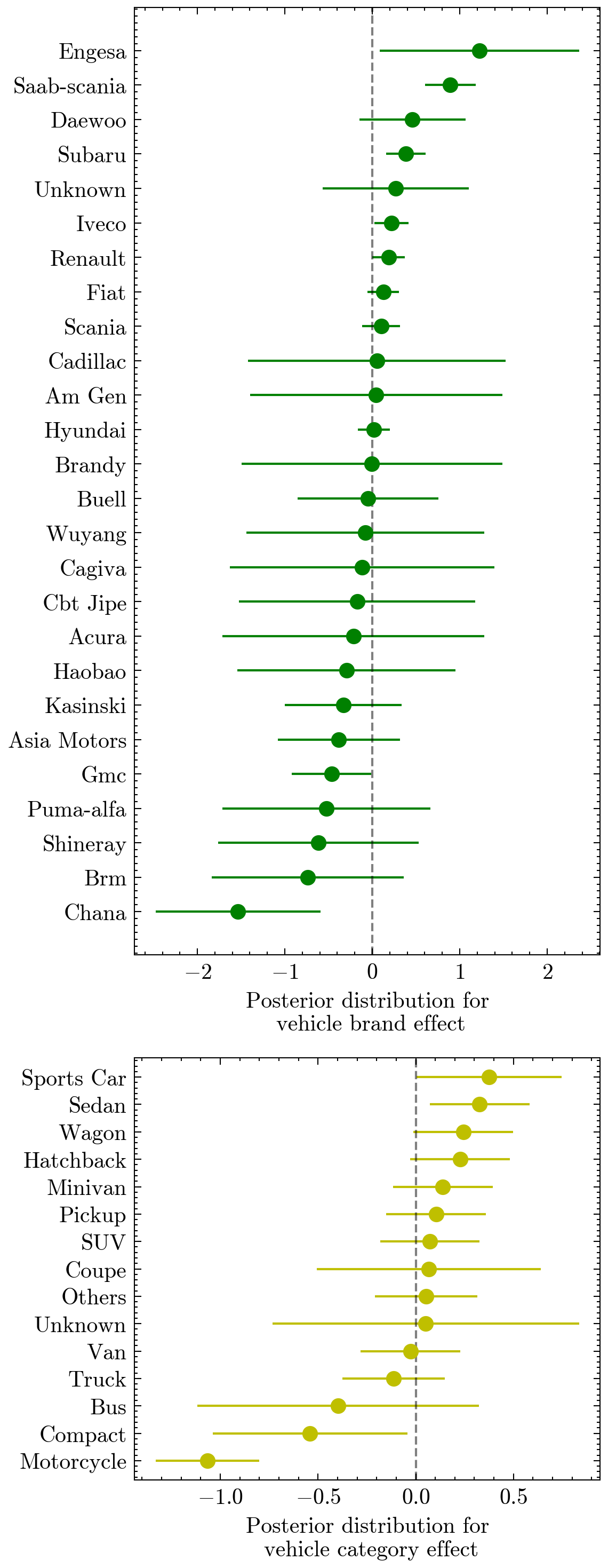}
\caption{Posterior summaries for parameters linked to vehicle brand and category. Due to space constraints, we show only every fourth brand. All categories are displayed. Lines indicate 2$\sigma$ credible interval bounds and dots represent posterior mean estimates of model parameters.}
\label{fig:category-brand}
\end{figure}

As demonstrated in Figure \ref{fig:category-brand}, the posterior estimates for the categorical predictors vary significantly, with some categories showing notable increases in expected claim counts and others decreases. Brands with large impacts include \emph{Engresa}, \emph{Subaru}, and \emph{Ducati}. Approximately one quarter of the brands showed posterior coefficient distributions markedly different from zero, as determined by 95\% credible intervals. In terms of vehicle type, sports cars tend to be associated with higher claim counts, whereas trucks, buses, compact cars, and motorcycles typically have lower claim counts. Nonetheless, only compact cars and motorcycles display significantly negative coefficients, according to 95\% credible intervals.

% EDIT 1: Removed "highly" before "variable" 
% EDIT 2: Changed "substantial" to "notable"
% EDIT 3: Changed "Roughly" to "Approximately"
% EDIT 4: Changed "exhibited" to "showed"
% EDIT 5: Changed "significantly" to "markedly"
% EDIT 6: Removed "perhaps unsurprisingly" (editorializing language)
% EDIT 7: Changed "show significantly" to "display significantly"
% EDIT 8: Changed "we only show" to "we show only"
% EDIT 9: Changed "are shown" to "are displayed"
% EDIT 10: Changed "indicate" to "represent" in the caption
\subsection*{Spatial effect}
\begin{figure}[h]
\centering
\includegraphics[width=0.6\textwidth]{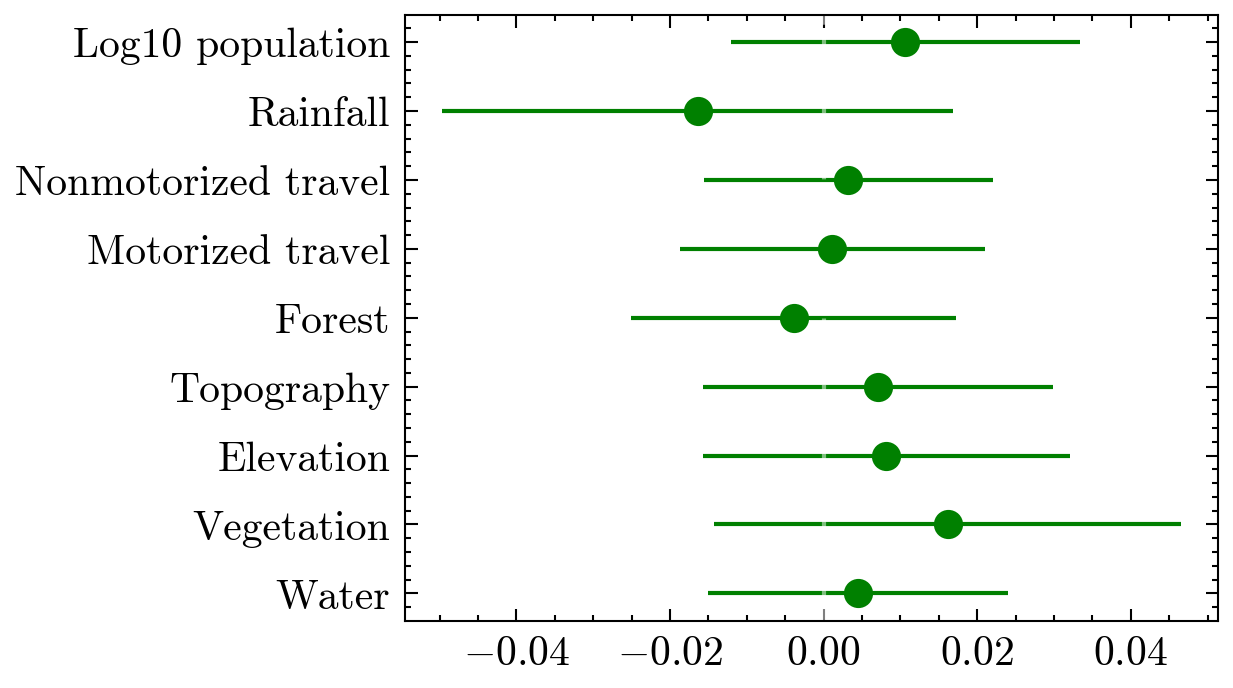}
\caption{Posterior summaries for parameters relating city-level covariates to predicted counts. Lines indicate 2$\sigma$ credible interval bounds and dots indicate posterior means.}
\label{fig:geo-coefs}
\end{figure}

\begin{figure}[h]
\centering
\includegraphics[width=0.8\textwidth]{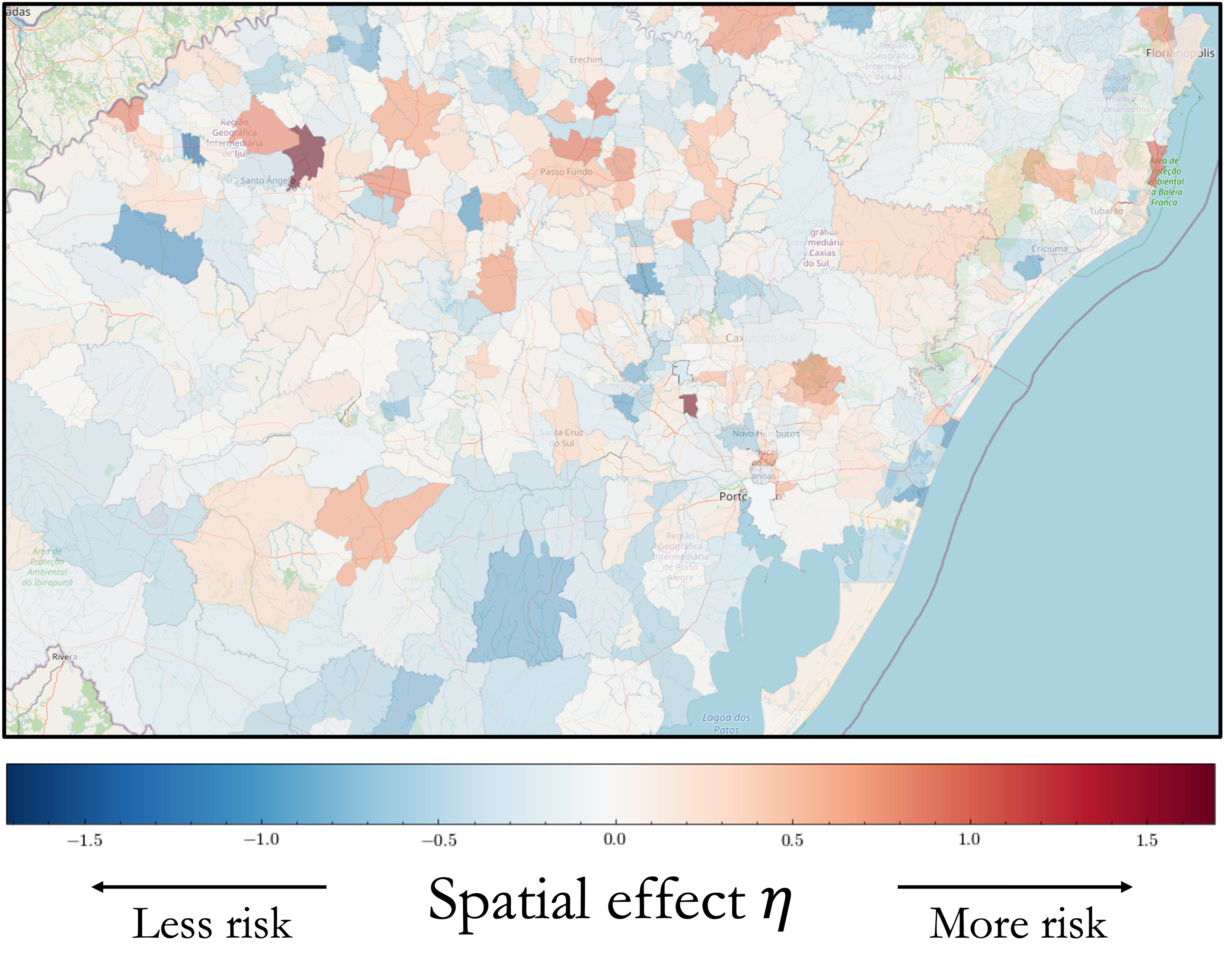}
\caption{Posterior mean estimates of spatially correlated values of $\eta$ in the state of Rio Grande do Sul.}
\label{fig:car}
\end{figure}

Despite the inclusion of several city-level covariates, we find that none of these covariates' coefficients have values significantly different from zero, assessed via the 95\% posterior credible intervals (Figure \ref{fig:geo-coefs}). The posterior mean values for these coefficients suggest % EDIT 1
that population density and vegetation may be % EDIT 2
positively associated with claim counts, while rainfall and forest cover appear to be % EDIT 3
negatively associated with claim counts. % EDIT 4
There is also evidence of spatial autocorrelation in $\bm{\eta}$ values, demonstrating the utility of this model component in identifying patterns of risk which are not fully explained by covariates, as depicted in Figure \ref{fig:car}.

% EDIT 1: Replace "exhibit" with "have".
% EDIT 2: Replace "provide weak evidence" with "suggest".
% EDIT 3: Add "may be" to maintain non-assertive language.
% EDIT 4: Replace "anticorrelated" with "negatively associated".
\subsection*{Temporal effect}
\begin{figure}[h]
    \centering
    \includegraphics[width=0.6\textwidth]{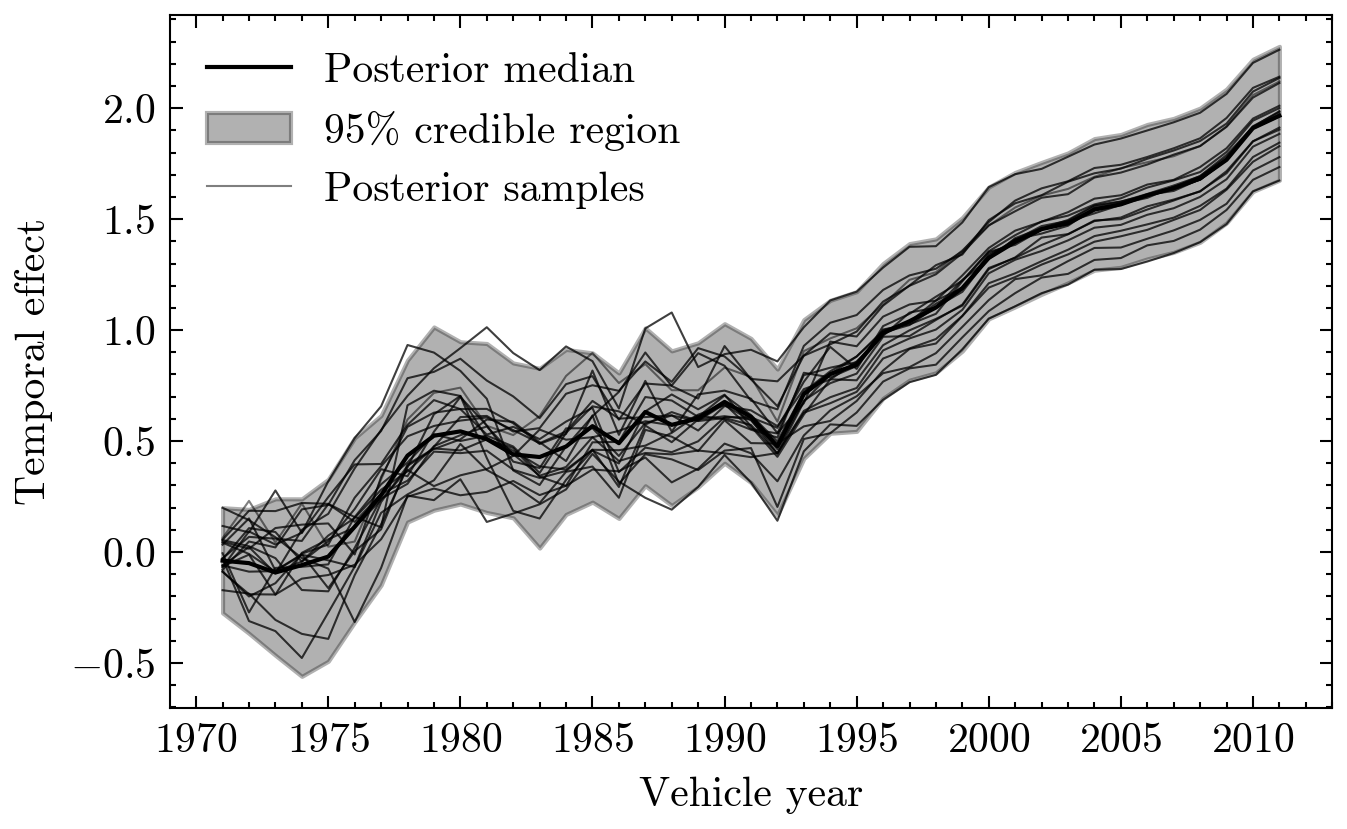}
    \caption{Posterior summary for time-varying effect.}
    \label{fig:grw}
\end{figure}

The impact of car manufacture date on expected collision claim counts is shown in Figure \ref{fig:grw}. The relationship between car age and claim count is relatively level for vehicles aged 20 to 30 years. We also noted a nearly linear increase in this effect for cars younger than 20 years, with the trend intensifying as vehicle age decreases. In summary, the risk assessed is stable for vehicles ranging from 20 to 30 years old and decreases with the age for cars under 20 years old.

% EDIT 1: Changed "captured by" to "shown in"
% EDIT 2: Changed "flat" to "level"
% EDIT 3: Removed "assessed"
% EDIT 4: Changed "inversely proportional" to "decreases with"
\section{Discussion}
The case study presented in this work has integrated several modeling subcomponents to address different aspects of risk modeling for insurance. We note while elements such as the spline-based exposure adjustment, vehicle category, and time-varying effect demonstrated strong signals manifesting as coefficients or parameters with posterior credible intervals excluding zero, the city-level covariates did not. We also observed a relatively large number of model parameters and coefficients with wide posterior credible intervals crossing zero, suggesting that the data is not strong enough to constrain these values to useful ranges. A major limitation of this work is the absence of shrinkage priors, which are designed specifically for sparsification \citep{carvalho09}, preventing us from applying substantial regularization to push parameter estimates towards zero when needed. Employing such priors would also allow for more model components, like feature-brand interactions or varying effects for the categorical predictors. A more comprehensive analysis could include policy-level covariates like driver age, driving record, and other risk or exposure indicators. This was intended as a demonstration for insurance-relevant models using large ($>1000$) parameter and data sets in a general-purpose statistical framework like NumPyro. Nonetheless, there is no barrier to creating models with significantly more parameters and data, given enough computational resources. We anticipate future studies using multi-GPU setups, enabling models with $10^5$ or more parameters and $>10^8$ observations.

% EDIT 1: Changed "case study shown" to "case study presented"
% EDIT 2: Removed "distinct" before "aspects"
% EDIT 3: Shortened "showed strong signal manifesting as" to "demonstrated strong signals manifesting as"
% EDIT 4: Removed "this" before "suggesting"
% EDIT 5: Changed "by avoiding" to "the absence of"
% EDIT 6: Changed "to push" to "to push towards"
% EDIT 7: Removed "potentially" before "allow for"
% EDIT 8: Removed "also potentially" and changed to "also allow for"
% EDIT 9: Changed "richer analysis might" to "more comprehensive analysis could"
% EDIT 10: Changed "like" to "such as" before "driver age"
% EDIT 11: Changed "look forward to future" to "anticipate future"
% EDIT 12: Added "enabling" before "models with $10^5$"
\subsection*{Future work}
We anticipate further interest in statistical modeling for insurance with large datasets and substantial models, and we view several avenues as particularly relevant directions for future work. There exists considerable potential for extending the model to include multiple types of counts with interesting correlation structures \citep{bermudez11} to help share information from different observation types \citep{anastasiadis12}. For example, we can imagine a scenario where the risk due to robbery and the risk due to fire are highly correlated because of underlying factors related to emergency services and population density. Individual models of each type may be fit on data without a strong enough signal to constrain parameter estimates; together, however, a multivariate model may share information and consequently enable more insights into joint risk factors. To accommodate data with varying degrees of dispersion reflecting different amounts of underlying heterogeneity, using likelihoods suitable for multiple dispersion types \cite{guikema08} may be productive. For a more practical and useful modeling exercise of direct relevance to insurance operations, we may want to model claim frequency and severity together \citep{frees08, chin23} since both types of outcomes are essential for pricing and reserving.

In early modeling iterations, we identified prominent disparities in risk between municipalities close to major urban centers and those farther away. While the model version ultimately presented explains a significant amount of this variation using covariates and the spatial random effect, we are interested in using a latent mixture model to cluster together cities with similar excess risk patterns. Generally, this type of model is hard to accommodate in a gradient-based MCMC framework as the latent class label is usually discrete and thus does not admit a log-posterior which is differentiable in all latent variables. However, both Stan and NumPyro have the ability to either enumerate or automatically marginalize out latent variables of modest cardinality, thus avoiding this issue and allowing for the use of gradient-based MCMC. We would also be interested in using computationally less expensive approximations to the Poisson likelihood \citep{cameron97} to speed up model fitting.

% EDIT 1: Changed "rich models" to "substantial models"
% EDIT 2: Changed "several avenues" to "particularly relevant directions"
% EDIT 3: Changed "correlation structure" to "correlation structures"
% EDIT 4: Removed "highly" as it was editorializing 
% EDIT 5: Changed "a strong" to "a strong enough"
% EDIT 6: Changed "fruitful" to "productive"
% EDIT 7: Changed "may wish" to "may want"
% EDIT 8: Changed "and thereby allow for" to "and consequently enable more"
% EDIT 9: Changed "using" to "suitable for"
% EDIT 10: Removed "necessarily" as it was unnecessary
% EDIT 11: Changed "prominent" to "strong"
% EDIT 12: Changed "great deal" to "significant amount"
% EDIT 13: Changed "challenging" to "hard"
% EDIT 14: Changed "cheaper" to "less expensive"
% EDIT 15: Changed "accelerating" to "speed up"
\section{Conclusion}
This study demonstrates the value of state-of-the-art statistical modeling frameworks like NumPyro that work seamlessly on GPUs to speed up model fitting without sacrificing any flexibility in the selection of model components. We developed a log-additive model for collision claim frequencies of auto insurance policies in Brazil. This model includes a nonlinear exposure adjustment, categorical predictors, a city-level random effect with spatial correlation, and a dynamic temporal effect. The model succeeded in explaining much of the data variance, though the city-level covariates showed weak effects. Moreover, we discovered that NumPyro's GPU-based implementation made the model fitting process nine times faster than using a CPU, as shown in the log posterior gradient calculations, enabling completion in a practical timeframe. This study should interest both practitioners and researchers in the insurance sector, as well as those who employ Bayesian methods for analyzing extensive spatial datasets.

% EDIT 1: Changed "highlights the utility of" to "demonstrates the value of"
% EDIT 2: Changed "general-purpose" to "state-of-the-art" and removed "accelerated" before "model fitting"
% EDIT 3: Changed "constructed" to "developed"
% EDIT 4: Changed "auto insurance policies in Brazil" to "auto insurance policies in Brazil" for clarity
% EDIT 5: Changed "claim counts" to "claim frequencies" for precision
% EDIT 6: Removed "a great deal of" as editorializing and replaced with "much of"
% EDIT 7: Changed "did not exhibit strong signal" to "showed weak effects"
% EDIT 8: Changed "provided a 9x speedup" to "made the model fitting process nine times faster" for clarity
% EDIT 9: Changed "researchers and practitioners" to "practitioners and researchers" to match the subject order in the following phrase
% EDIT 10: Changed "with large, spatial data sets" to "for analyzing extensive spatial datasets" for smooth reading
\section{Data Availability} % EDIT 1: Capitalized "Availability"
All code and data are available for replication at \url{https://github.com/ckrapu/bayes-at-scale}. % EDIT 2: Changed "reproduction" to "replication"

\newpage
\bibliography{references}

\end{document}